# Self-Injection Locking and Phase-Locked States
# in Microresonator-Based Optical Frequency Combs

Pascal Del'Haye[1*], Scott B. Papp[1], Scott A. Diddams[1]
[1]*National Institute of Standards and Technology (NIST), Boulder, CO 80305, USA*

**Microresonator-based optical frequency combs have been a topic of extensive research during the last few years. Several theoretical models for the comb generation have been proposed; however, they do not comprehensively address experimental results that show a variety of independent comb generation mechanisms. Here, we present frequency-domain experiments that illuminate the transition of microcombs into phase-locked states, which show characteristics of injection locking between ensembles of comb modes. In addition, we demonstrate the existence of equidistant optical frequency combs that are phase stable but with non-deterministic phase relationships between individual comb modes.**

Optical frequency combs have been a transformational physical measurement tool since their inception more than a decade ago [1, 2]. As a ruler for optical frequencies, they enable precision measurements in the fields of optical clocks [3, 4], astrophysical spectrometer calibration [5], and spectroscopy [6-8]. On the applied side, optical frequency combs are promising tools for multi-channel generators in telecommunications [9, 10], gas sensing [11], as optical and microwave frequency references [12], and for arbitrary optical waveform generation [13, 14]. To date most optical frequency combs are based on femtosecond mode-locked lasers [2]. However, during the past five years a novel type of comb generator [15-35] has sparked significant scientific interest in new techniques and expanding applications of optical comb generation. The new comb generation principle is based on parametric four-wave mixing in a monolithic high-Q micro-resonator and does not utilize conventional stimulated laser emission. However, in contrast to a mode-locked laser comb there is little consensus on possible mode-locking mechanisms (e.g. a saturable absorber) in microcombs, which could act to enforce equidistance of the modes and align the phases to generate ultrashort optical pulses. This is the case for both experimental research on microcombs and recent theoretical studies that focus on a better understanding of the actual comb generation principle [36-44]. In this work, we present an extensive frequency-domain analysis on the transition of microcombs into phase-locked states. We show the presence of a self-injection locking mechanism within the resonator that is mediated via the parametric gain, and support our measurements with a model for self-injection locking of microcombs. In addition, we present phase measurements of the comb modes and show the existence of phase stable microcombs with uniform mode spacing, but a non-deterministic phase relationship between individual comb lines. In contrast to mode-locked lasers and recent observation of soliton generation in a microcomb [45], these measurements indicate the existence of phase-locked microcombs without a circulating high-peak-power pulse and related conventional mode-locking mechanisms.

For our experiments, the microcomb is generated from a tunable diode laser that is amplified and coupled via a tapered optical fiber into a whispering-gallery mode of a fused silica microrod resonator (diameter 2.6 mm, mode spacing ~25.6 GHz, loaded quality factor $Q = 1.85 \times 10^8$) [34, 46]. When the laser is tuned into a cavity resonance from the blue side, the resonator locks itself thermally to the laser [47] and generates an optical frequency comb. For resonators like the one we employ, the dispersion [48] dictates that the parametric gain is maximum for modes that are not adjacent to the pump. Just above threshold, this can lead to clustered or bunched combs (Figure 1a), with bunches having the same mode spacing but mutual offsets [41, 49], as illustrated in Figure 1b. Clearly, such offsets represent a departure from the uniform nature of an ideal frequency comb; however, we observe that behavior akin to injection-locking can occur as bunches begin to overlap, leading to an offset-free, continuous comb. Similar phase-locking effects have already been indicated by disappearing sidebands in the microwave mode spacing beat note of microcombs [28].

In order to analyze the phase-locking behavior of a microcomb system and confirm the existence of a uniform and continuous comb, we measure the frequencies of the microcomb modes relative to those of an established mode-locked laser comb (see Fig. 1c). This is accomplished with a multi-channel synchronous phase/frequency recorder that measures the frequency spacing between the pump laser and n-th comb sideband $\Delta f_n$ as well as the microcomb's mode spacing $f_{mc}$. Any possible offset of the n-th mode from its expected position is given as $f_{off} = n \times f_{mc} - \Delta f_n$. The microcomb spacing $f_{mc} = f_{mw} + f_1$ is measured by sending the comb light onto a fast photodiode, mixing the signal down with a microwave reference $f_{mw}$ and recording frequency $f_1$. The value $\Delta f_n = m \times f_3 \pm f_4 \pm f_2$ is measured with the reference comb by counting its repetition rate as well as beat notes with the pump laser and the n-th sideband, respectively (cf. Fig. 1d). In this way we obtain the offset as $f_{off} = n \times (f_{mw} + f_1) - m \times f_3 \pm f_4 \pm f_2$. The ambiguity of the $\pm$ signs in $f_{off}$ can be resolved by slightly changing the offset of the reference comb and monitoring whether the beat frequencies $f_2$ and $f_4$ increase or decrease.

The experimental setup in Fig. 1 allows us to monitor offsets in microcombs in real time (only limited by the averaging time of the synchronous phase/frequency recorder). For this measurement, we tune a microcomb into a state close to zero offset by changing the coupling to the resonator and the detuning between pump laser and resonator mode. Once in this state, the offset can be fine-tuned by slightly changing the power launched into the resonator (Fig. 2a). Here, the x-axis if offset by ~50 mW, which corresponds to the average launched power. It is observed that the measured offset locks to zero in a characteristic way that is known from injection locking [50, 51]. This locking behavior can be modeled from the Adler equation that describes the phase evolution for an injection locked signal:

$$\frac{1}{2\pi} \frac{d\varphi(t)}{dt} = \Delta\nu - \frac{\nu_0}{Q} \frac{E_1}{E_0} \sin \varphi(t) \ . \qquad (1)$$

Here, $\nu_0 \approx 193$ THz is the oscillator's free-running frequency, $Q = 1.85 \times 10^8$ is the loaded quality factor of the resonator, and $E_1/E_0$ is the relative field amplitude of the injected and free-running frequencies (in the case of a microcomb, this corresponds to the combined relative amplitudes of the comb modes in the regions of overlapping bunches). The variable $\Delta\nu = \nu_0 - \nu_1$ corresponds to the difference between injected frequency $\nu_1$ and free-running oscillator frequency $\nu_0$. In other words, $\Delta\nu$ is the offset in a hypothetical microcomb where no locking mechanism is present. In the case of self-injection locking in a microcomb, $\nu_0$ and $\nu_1$ are not directly accessible; however, the difference $\Delta \nu$ tunes linearly with the launched power change $\Delta P$ such that $\Delta \nu = \kappa \times \Delta P$, with $\kappa$ being a constant.

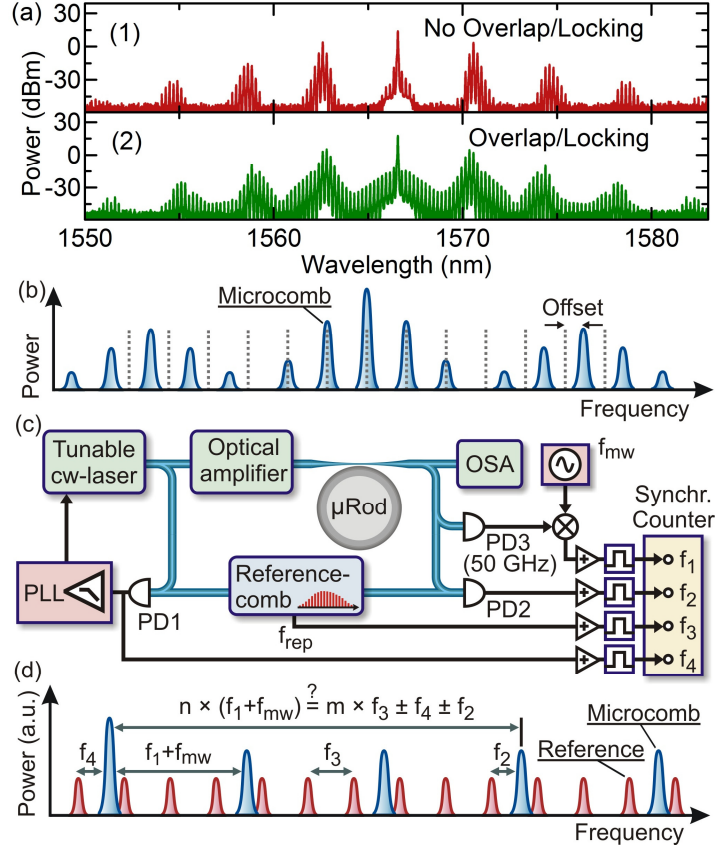

FIG. 1. (a) Optical spectra of microcombs without (1) and with (2) overlap. The amount of overlap can be controlled by changing the detuning of the pump laser with respect to the resonator mode. (b) Offset in a bunched microcomb. The dashed lines depict the position of equidistant comb modes. (c) Experimental setup to analyze and test self-injection-locking and phase-locking behavior of microcombs. PLL=phase-locked loop, PD=photodiode. (d) Scheme to determine whether the frequency of a comb mode is commensurate with that of a uniformly-spaced comb. The spacing between pump mode and the n-th sideband is measured with a reference comb and compared to the expected spacing given by n-times the microcombs' mode spacing.

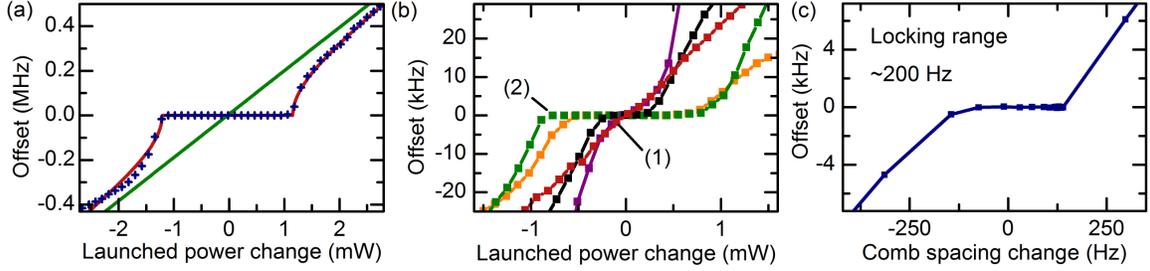

FIG. 2. Injection locking of bunched comb states. (a) Offset of the 37$^{th}$ sideband of a bunched microcomb vs. launched power. Injection locking is observed when the offset approaches zero (blue crosses). The red line is a fit based on the Adler equation for injection locking. The green curve is obtained from the fit and shows how the offset would tune without injection locking mechanism present. Panel (b) shows the injection locking behavior for different comb states with different overlap between bunches. Trace (1) shows a comb state without overlap between bunches, which shows no injection locking. Trace (2) shows a state with strong overlap that leads to injection locking. Corresponding optical spectra are shown in Fig. 1a. The overlap is controlled by changing the detuning between pump laser frequency and microresonator mode. Panel (c) shows the injection locking range as function of the microcomb spacing (tuned via launched power change).

The solutions of the Adler equation (1) are periodic waveforms with an average frequency that is given by (cf. Ref. [50])

$$f_{\text{off}} = \langle \frac{d\varphi(t)}{dt} \rangle = \Delta\nu \frac{\sqrt{K^2-1}}{|K|} = \kappa \times \Delta P \frac{\sqrt{K^2-1}}{|K|} \quad , \quad (2)$$

with

$$K = 2Q \frac{E_0}{E_1} \frac{\kappa \, \Delta P}{\nu_0} \quad . \quad (3)$$

A fit of the data in Fig. 2a using equation (2) shows excellent agreement with the injection locking model. The only free parameters for the fit are $\kappa = (195 \pm 2) \frac{\text{kHz}}{\text{mW}}$ and the ratio $E_0/E_1 = 2.27 \pm 0.03$ (corresponding to a power ratio of $P_0/P_1 = (E_0/E_1)^2 = 5.2 \pm 0.1$). The parameter $\kappa$ corresponds to the offset tuning for $|\Delta P| \gg 1$ (or for zero coupling between bunches). In this case the last term $\left(\sqrt{K^2-1}/|K|\right)$ in equation (2) approaches unity such that $f_{\text{off}} \approx \Delta \nu = \kappa \times \Delta P$ (green line in Fig. 2a). The microcomb self-injection locks for $K^2 \leq 1$, leading to a locking range of

$$\Delta \nu_{\text{lock}} = \kappa \, \Delta P_{\text{lock}} = \frac{\nu_0}{Q} \frac{E_1}{E_0} \quad . \quad (4)$$

In this range, the real part of $f_{\text{off}}$ vanishes, leaving only an imaginary part. This imaginary part corresponds to a phase shift $\Delta\phi$ between the (self-)injected frequency $\nu_1$ and $f_{\text{off}}$ with $-\frac{\pi}{2} < \Delta\phi < +\frac{\pi}{2}$. The other fitting parameter from equation (2) and (3) is the ratio $E_0/E_1$, which determines the width of the locking range. Using equation (4) we obtain $\Delta\nu_{\text{lock}} = (460 \pm 6)$ kHz for the locking range obtained from the microcomb data of Fig. 2a.

The importance of the overlap between bunches for self-injection locking is shown in Fig. 2b. The graph shows the locking region for different comb states with different overlap between bunches. In particular, no locking behavior is observed when the microcomb bunches are far from overlapped (traces (1) in Fig. 2b with corresponding optical spectrum in Fig. 1a). The amount of overlap between the comb bunches is controlled by changing the detuning of the pump laser frequency with respect to the microresonator mode. Fig. 2c shows the measured offset as function of the comb spacing change. Here, the comb spacing is changed again by tuning the launched power. We can see that the microcomb spacing is more stable within the injection- locked range and tunes by only ~200 Hz.

In the same resonator mode, but at smaller detunings of the pump laser with respect to the microresonator mode, we observe another phase-locking behavior that has different characteristics compared to the self-injection locking. In this case, we cannot tune continuously into the locked state, but the resonator rather "jumps" into a stable comb state with a characteristic and symmetric shape of the optical spectrum and low noise in the mode spacing beat note signal. One example of these comb states is shown in Fig. 3a. The same setup shown in Fig. 1c is used to determine whether the spectrum in Fig. 3a constitutes a continuous optical frequency comb without offsets. An optical filter is used in this measurement in order to select and count the offset of single comb sidebands. Fig. 3b shows the result of such an offset measurement for the 98$^{th}$ sideband of the microcomb. The measured offset shows a mean value of 740 $\mu$Hz $\pm$ 3 mHz, which is consistent with zero and limited by the measurement time. Several comb modes (marked with arrows in Fig. 3a) have been measured in this way and all of them reveal a measurement-time-limited offset consistent with zero. Fig. 3d is the Allan deviation in Hz for offset measurements of the 98$^{th}$ and 37$^{th}$ sideband. These data show that the measured offset averages down to a few millihertz (consistent with zero) at 1000 seconds measurement time. The Allan deviation is derived from the measurement in Fig 3b by grouping data points together (this is possible because the employed frequency counter has zero dead-time between measurements).

Further confirmation of the offset-free nature of this microcomb is shown in Fig. 1e where we observe the correlation between the microcomb mode spacing $f_{mc}$ and the frequencies of the 37$^{th}$, 61$^{st}$ and 98$^{th}$ microcomb modes. Since the microcomb is not actively locked, we simultaneously observe the drift in $f_{mc}$ and the sideband positions during the measurement. The measurements show that the measured sidebands tune exactly as expected as $n \times f_{mc}$ (with n being the sideband number). Linear fits of the data in Fig 3e (dashed lines) yield slopes of 97.995 $\pm$ 0.004, 60.99 $\pm$ 0.02, and 36.9998 $\pm$ 0.0004 for the 98$^{th}$, 61$^{st}$, and 37$^{th}$ sideband, respectively (with the uncertainty limited by the relatively short measurement time).

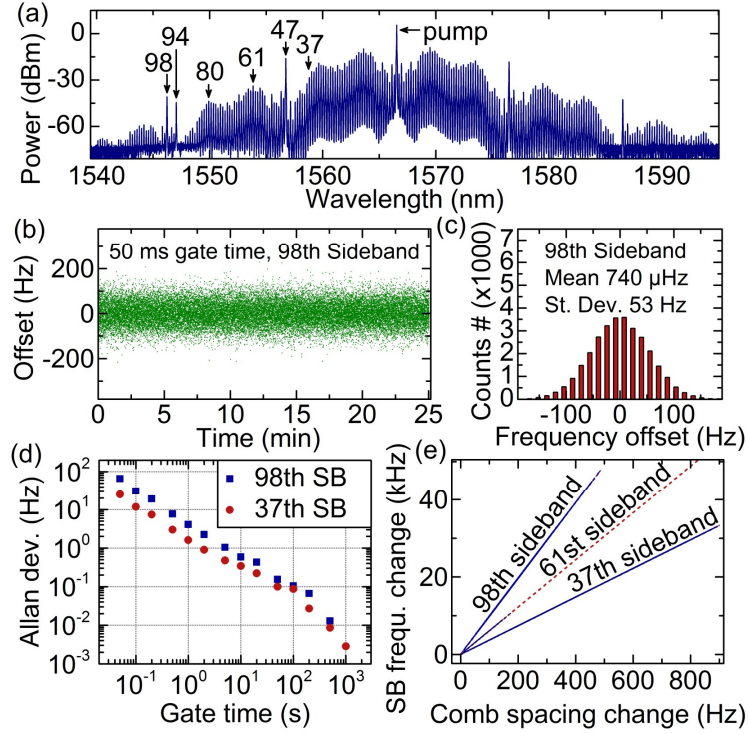

FIG. 3. (a) Optical spectrum of a phase-locked microresonator comb. The comb is generated in a 2.6-mm-diameter fused silica rod-resonator and consists of more than 200 individual lines. The comb's mode spacing is 25.6 GHz. (b) Measurement of the offset of the +98$^{th}$ comb sideband in a phase-locked state. The counter gate time for this measurement is 50 ms and the mean value of the whole measurement is 740 µHz. (c) Distribution of the data in panel (b). (d) Allan deviation (not normalized) of the offset of the +37$^{th}$ and +98$^{th}$ comb sideband. (e) Tuning of the 37$^{th}$, 61$^{st}$ and 98$^{th}$ sideband versus the measured comb spacing. The modes in this phase-locked state tune as expected by n-times the comb spacing.

It is interesting to note that these phase-locked comb states are remarkably stable and can run continuously for more than two days without any pump laser frequency stabilization or temperature stabilization. This stability permits more extensive measurements of the relative phases of the microcomb modes. In order to measure the phases of comb modes, we implement a liquid-crystal-based programmable "waveshaper" that enables control of the phases and amplitudes of individual microcomb modes. In conjunction with a nonlinear optical autocorrelator, this allows us to phase-align all the microcomb modes such that they generate a short pulse [14, 22]. Subsequently, we use this same control and measurement capability to infer the phase relationship among the modes within the microresonator. Note that by phase we mean the phase offset $\phi_n$ and not the time evolving instantaneous phase $(\omega_n t + \phi_n)$ of the n-th comb mode. The experimental setup for the phase and amplitude control is shown in Fig. 4a. Note that the dispersion of the setup between microresonator and autocorrelator has to be taken into account in this measurement. In order to calibrate the measurement and measure the dispersion of the setup, a pulsed laser is sent through the same setup and the waveshaper phases are adjusted in a way that the whole setup is dispersion-free. Fig. 4b and 4c show optical spectrum and autocorrelation of a phase-locked comb state without any phase adjustments. The autocorrelation has a background to peak ratio of ~0.5, which is characteristic for a comb spectrum with randomly-distributed phase offsets. Fig. 4d and 4e show the spectrum and autocorrelation of the same comb after amplitude flattening and phase optimization. The pulse length of the autocorrelation is Fourier-limited with a width of ~290 fs assuming a sinc-pulse shape (which is expected from a rectangular shaped optical spectrum). Once the static amplitude and phase mask is applied, this pulse form is stable as long as the microcomb remains in the same phase-locked state.

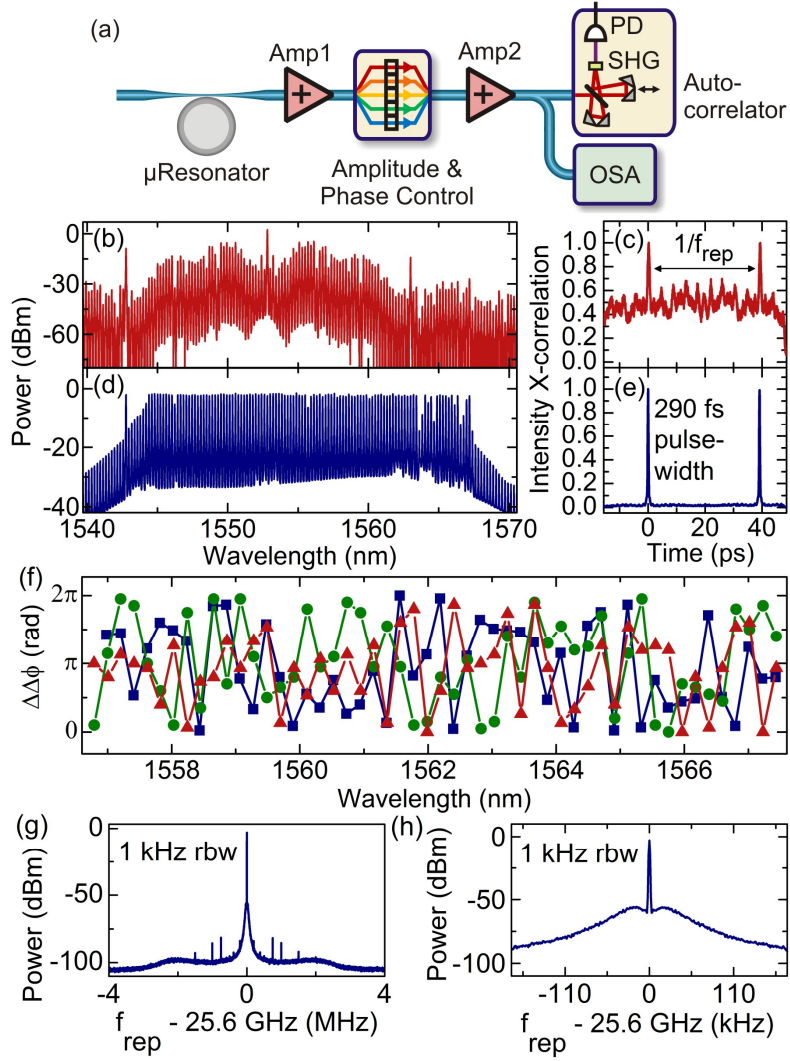

FIG. 4. Phase measurement of microresonator modes via optical autocorrelation. (a) Experimental setup for phase measurements. A liquid crystal array-based waveshaper is used to adjust the amplitude and phase of individual microcomb modes. The phases are optimized based on the peak power signal from an optical autocorrelator. (b),(c) Optical spectrum and autocorrelation without amplitude and phase adjustments. (d), (e) Spectrum and autocorrelation after phase adjustment. (f) Second-order dispersion of the phases of three different phase-locked states. The phases of the comb modes in all three states appear to be random and unique, but are stable in time over more than 48 hours. (g),(h) Mode spacing beat note of a phase-locked microcomb at different scales.

Knowledge of the phases that have been applied to the comb modes to generate a short pulse combined with the additional measurement of the setup dispersion allows us to calculate the phases of the modes at the point where the comb exits the resonator. It is important to know that the resulting pulse shape is independent of a constant offset of all phases and also independent of a linear increase in phase with frequency (the latter just shifts the pulse in time). Thus, in order to eliminate linear phase changes, we analyze the second derivative of the phases, which we define as the phase dispersion

$$\Delta\Delta\phi \equiv \phi_{n+1} + \phi_{n-1} - 2\phi_n \quad . \quad (5)$$

In case of quadratic dispersion in the comb modes, $\Delta\Delta\phi$ is expected to be constant. Taking third order dispersion into account, $\Delta\Delta\phi$ would increase linearly. Figure 4f shows measurements of $\Delta\Delta\phi$ for three different phase-locked comb states. We find that there is no evident pattern in the distribution of $\Delta\Delta\phi$ and moreover, the phase dispersion is different for all measured phase-locked states. Running the phase retrieval twice for the same phase-locked state returns the same distribution for $\Delta\Delta\phi$ with a mean absolute deviation of $\sim 2\pi/13$ rad between the results, which is a good estimate for the error on the phase measurement. As implied from the stable time-domain waveform, the measured phase dispersion was verified to be constant for a given phase-locked state over more than 48 hours; despite pump laser frequency drifts of several 100 MHz and lab temperature changes, which affect the coupling into the resonator. Together with the data in Fig 3, this measurement indicates the existence of stable phase-locked microcomb states with non-deterministic phase distributions among the comb modes.

Finally, Figures 4g,h show the mode spacing beat note of an amplitude- and phase-optimized frequency comb at ~25.6 GHz. The signal-to-noise ratio of more than 100 dB in 1 kHz resolution bandwidth is further evidence of stable phase-locked operation.

In conclusion, we have shown the existence of self-injection locking of optical frequency combs in microresonators. Even though the injection locking takes place simultaneously between many different comb modes, it is possible to describe the process as a simple two-frequency injection locking mechanism. Moreover, we have analyzed intrinsically phase-locked states in a microresonator with apparently disorganized, but stable phases in time. These results are an important step towards a better understanding of different mechanisms responsible for optical frequency comb generation in monolithic microresonators. In particular, they imply that injection-locking between multiple parametric processes may play a role in stable parametric comb generation. Additionally, our results demonstrate the existence of frequency comb generation in microresonators that does not involve mode-locking mechanisms favoring high-peak power. This is a surprising departure from the processes required in conventional mode-locked lasers. Moreover, we have shown that the generated combs are long-term phase stable and can be transformed into short pulses, which is an important pre-requisite for spectral broadening and self-referencing. In future research it will be important to understand how the presented self-injection locked and phase-locked states are related to each other and whether there is a connection to previously reported time-domain soliton generation [45] in microresonator-based optical frequency combs.

**Acknowledgements:** This work is supported by NIST, the DARPA QuASAR program, and NASA. PD thanks the Humboldt Foundation for support. We further thank Frank Quinlan and Jeff Sherman for helpful comments on the manuscript. This paper is a contribution of NIST and is not subject to copyright in the United States.

[*]pascal.delhaye@gmx.de